\documentclass[%
 reprint,
 amsmath,amssymb,
 aps,
floatfix
]{revtex4-2}
\usepackage[autostyle]{csquotes}
\usepackage{graphicx}
\usepackage{dcolumn}
\usepackage{bm}
\usepackage{hyperref}
\usepackage{gensymb}
\usepackage{amsmath}
\hypersetup{colorlinks = True,
linkcolor = blue,
filecolor = magenta,
urlcolor = blue,
citecolor = blue}

\begin{document}

\preprint{APS/123-QED}

\title{Generation and certification of pure phase entangled light}

\author{Rounak Chatterjee}
\author{Mayuresh Kanagal}
    \author{Vikas S Bhat}
    \author{Kiran Bajar} 
    \author{Sushil Mujumdar}
     \email{mujumdar@tifr.res.in; http://www.tifr.res.in/~nomol}
    \affiliation{%
     Tata Institute of Fundamental Research, 400005 Mumbai, India
    }%

\date{\today}

\begin{abstract}
Biphoton systems exhibiting entanglement in position-momentum variables, known as spatial entanglement, are among the most intriguing and well-studied phenomena in quantum optics. A notable subset of these are \textit{phase entangled states}, where entanglement manifests purely through correlations in the spatial phase of the wavefunction. While the generation of such states from biphotons via spontaneous parametric down-conversion has been explored, their physical implications and applications remain under-investigated. In this work, we theoretically and experimentally examine a unique form of phase entanglement known as `pure' phase entanglement. This state exhibits the unusual feature that the position of one photon is correlated with the momentum of the other. Unlike typical spatially entangled states, it shows no direct correlation in position or momentum between the two photons, underscoring that all correlations arise purely from the spatial phase of the wavefunction. We delve deeper into the theory of this state and experimentally construct it from known phase-entangled states. To certify its properties, we propose a setup that performs a “one-particle momentum measurement” and explore the various tunable parameters. We also highlight potential applications of this state in quantum optics and imaging experiments.
\end{abstract}
\maketitle

\section{Introduction}
\label{sec:Introduction}
Generation of new and exotic entangled states \cite{Horodecki2009} has consistently offered researchers novel perspectives into quantum systems at an unprecedented level. These states are not only academically intriguing—enabling deeper exploration of the underlying physics—but also open up technological and application frontiers inaccessible through classical means. Prominent examples include the well-known EPR states \cite{Einstein1935}, bipartite entangled states \cite{Aspect1982}, and more complex forms such as GHZ states \cite{Greenberger1990} and NOON states. These states play crucial roles in quantum technologies, including quantum metrology \cite{Polino2020} and super-resolution imaging \cite{Shin2011}. The evolution of entangled state generation has thus been both profound and impactful.

\par One of the most well-studied methods to generate and investigate bipartite entangled systems involves time-correlated photon pairs produced via spontaneous parametric down-conversion (SPDC) in non-linear crystals. The resulting entanglement can manifest in various forms, such as polarization entanglement \cite{Kwiat1995}, orbital angular momentum entanglement \cite{mair_entanglement_2001,Krenn2017}, and spatial entanglement \cite{Howell2004,Moreau2012}. These systems have played critical roles in landmark experiments such as the Hong-Ou-Mandel effect \cite{HOM1987}, induced coherence \cite{Wang1991}, and quantum teleportation \cite{Bouwmeester1997}. Such phenomena underpin essential aspects of quantum information science and support emerging commercial technologies, including quantum communication \cite{gisin_quantum_2002}, dense coding \cite{mattle_dense_1996}, quantum repeaters \cite{Azuma2023}, and quantum imaging \cite{lemos_quantum_2014}.

\par In recent experimental efforts, researchers have explored entanglement in higher-dimensional degrees of freedom for applications including studies in disordered media \cite{valencia_unscrambling_2020} and high-dimensional quantum key distribution (QKD) \cite{Andrew2013}. However, perhaps the most academically pursued domain in bipartite systems is spatial entanglement \cite{howell2016,Walborn2010}. This form yields a diverse set of bipartite quantum states entangled in position-momentum variables and can be tailored for specific applications including precise fluorescence lifetime measurement\cite{Lyons2023}, light transmission through complex media \cite{Peeters2010,Devaux2023,lib_quantum_2022} and customized quantum computations \cite{Hugo2025} and high dimensional quantum cryptography\cite{Walborn2006_HDQKD,Vikas2025_HDQKD}, as well as in quantum imaging \cite{Ndagano2022,Defienne2024,zhang_background_2024}.

\par A defining feature of spatial entanglement in SPDC-generated bipartite systems is its manifestation as (anti-)correlations in position (momentum) observed in the near (far)-field regions of the two-photon transverse field \cite{Edgar2012}. While most studies and applications focus on these correlation properties, the underlying quantum state is capable of supporting far more exotic forms, which can be engineered using simple operations such as propagation or the addition of quadratic phases to the electromagnetic field. One particularly intriguing state that emerges via free-space propagation is the so-called “phase-entangled” state, where the wavefunction manifests its entanglement in the quantum phase rather than the amplitude. This non-intuitive state, although explored theoretically \cite{chan_transverse_2007,Tasca2009}, has rarely been studied experimentally \cite{just_transverse_2013,reichert_quality_2017,Rounak2025}. We believe that its counter-intuitive nature holds promise for revealing new directions in the study of continuous-variable entanglement. In this work, we aim to harness this non-intuitiveness to generate a novel quantum state with correlation properties previously unobserved in spatially entangled systems.

\par Mentioned briefly in \cite{chan_observable_2004}, the concept of a \textit{pure phase-entangled} state refers—at least theoretically—to a quantum state in which the two-photon spatial phase is entirely correlated, with no residual single-photon features. This stands in contrast to conventional phase-entangled light generated via SPDC \cite{chan_transverse_2007}, where residual single-photon characteristics typically persist. What makes this state especially compelling, in terms of its physical implications, is its unusual correlation structure: the position of one photon is tightly correlated with the momentum of the other. Unlike previously studied states—featuring position-position correlations in the near field or momentum-momentum anti-correlations in the far field \cite{Howell2004}—this hybrid correlation is unique. In earlier works introducing the concept of phase entanglement \cite{chan_transverse_2007}, the wavefunction showed no correlations in either position or momentum individually, but never both simultaneously. In contrast, the pure phase-entangled state exhibits complete absence of position-position and momentum-momentum correlations between the two photons, while maintaining a strong cross-correlation between the position of one and the momentum of the other. In this work, we explore the theoretical underpinnings of this state in depth and present an experimental protocol to generate it from position-correlated light. We also develop a measurement scheme to verify its unconventional properties.

\section{Theoretical background}
\label{sec:theory}
Phase-entangled states are characterized by the manifestation of entanglement through phase correlations, which become apparent depending on the basis—position or momentum—in which the quantum state is represented. Typically, the conditions under which a state exhibits phase entanglement in one basis do not coincide with those for its complementary basis, as clearly demonstrated in \cite{Tasca2009}. However, as discussed in sec.\ref{sec:Introduction}, a pure phase-entangled state exhibits phase correlations irrespective of the chosen basis. This invariance is evident from the generalized form of a pure phase-entangled state \cite{chan_observable_2004}, expressed in the position $(\psi_{p^2})$ and momentum $(\phi_{p^2})$ representations as:

\begin{equation}
\label{eqn:purephase}
\small \psi_{p^2}(x_1,x_2) = \sqrt{\frac{2A}{\pi}}\exp{\left(-A\left(x_1^2+x_2^2\right)-iBx_1x_2\right)}
\end{equation}
\begin{equation*}
\small \phi_{p^2}(p_1,p_2) \propto \exp{\left(-\bar{A}\left(p_1^2+p_2^2\right)-i\bar{B}p_1p_2\right)}
\end{equation*}

Here, $A$ and $B$ have units of inverse length squared, and the transformed parameters are given by $\bar{A} = A/(\hbar^2(4A^2+B^2))$, $\bar{B} = B/(\hbar^2(4A^2+B^2))$. The term “pure phase entangled” is thus justified either by the basis-independent nature of the phase correlations or, equivalently, by the fact that the phase contains only cross terms—$x_1 x_2$ or $p_1 p_2$—with no single-particle contributions. This strict phase correlation leads to a distinctive hybrid feature: the position of one photon is correlated with the momentum of the other. This feature becomes apparent upon performing a partial Fourier transform (\textit{PFT}), or a one-particle Fourier transform, which projects the quantum state into a mixed representation—position for one particle and momentum for the other:

\begin{eqnarray}
\label{eqn:PFT}
& \bar{\psi}_{p^2}(p_1,x_2) = \frac{1}{\hbar}\int{e^{-i\frac{p_1 }{\hbar}x_1}~\psi_{p^2}(x_1,x_2)~dx_1} \nonumber \\
& =\frac{1}{\hbar\sqrt{\pi}}\exp{(-Ax_2^2)}~\exp{\left(-\frac{\left(p_1+\hbar Bx_2\right)^2}{4\hbar^2A}\right)}
\end{eqnarray}

The conditional probability distribution derived from this wavefunction reveals that the conditional momentum of one photon, given the position of the other, follows a Gaussian distribution with conditional mean $\langle p_1|x_2\rangle = -\hbar Bx_2$ and standard deviation $\sigma_{p_1|x_2} = \hbar \sqrt{A}$. A more detailed physical interpretation will be provided in sec.\ref{sec:discussion}.

\par At first glance, this quantum state may appear to be purely theoretical, with no experimental counterpart. However, it is indeed possible to design an experimental setup that generates this state from the known phase-entangled state produced by spatially entangled light generated via spontaneous parametric down-conversion (SPDC) in a non-linear crystal. As established in SPDC theory \cite{Walborn2010,howell2016}, a specially cut non-linear crystal of length $L$, pumped by a narrow-bandwidth collimated laser (wavelength $\lambda_p$, refractive index in the crystal $n_p$, and field profile $E_p(\textbf{r})$), produces near-degenerate photon pairs at approximately twice the wavelength. These photon pairs are spatially entangled in both position and momentum, and can be engineered to propagate co-linearly with the pump beam. The wavefunction for such a state in the momentum variables of the two photons $(\textbf{p}_1, \textbf{p}_2)$ can be written as:

\begin{equation}
\label{eqn:SPDC original}
\psi(\textbf{p}_1,\textbf{p}_2) = \widetilde{E}_p(\textbf{p}_1+\textbf{p}_2) \text{ sinc}\left(\frac{L\lambda_p}{8\pi n_p}|\textbf{p}_1-\textbf{p}_2|^2\right)
\end{equation}

Here, $\widetilde{E}_p$ is the Fourier transform of the pump field $E_p$. Assuming the pump is an azimuthally symmetric Gaussian beam with waist $\sigma_p$ located at the center of the crystal, the above wavefunction can be approximated as a double Gaussian (DG) form \cite{law2004,Monken1998}. Fourier transforming this wavefunction into position space preserves the DG form. Due to system symmetry, the wavefunction in one-dimensional position variables becomes:

\begin{equation}
\label{eqn:DG}
\small \psi(x_1,x_2) =  \frac{1}{\sqrt{\pi \sigma_-\sigma_+}}\exp{\left(-\frac{(x_1-x_2)^2}{4\sigma_-^2}-\frac{(x_1+x_2)^2}{4\sigma_+^2}\right)}
\end{equation}

Here, $\sigma_- = \sqrt{\frac{L\lambda_p}{6\pi n_p}}$ and $\sigma_+ = \frac{\sigma_p}{2}$. These widths encode the degree of position correlation and momentum anti-correlation \cite{Edgar2012,Moreau2012}, as well as the overall entanglement content—quantified analytically by the Schmidt number, $K = \frac{1}{4}\left(\frac{\sigma_+}{\sigma_-} + \frac{\sigma_-}{\sigma_+}\right)^2$ \cite{law2004}. The parameters of the joint probability distribution (JPD), given by $|\psi(x_1, x_2)|^2$, can be measured via photon counting statistics using spatially pixelated photon detectors \cite{Edgar2012,defienne_general_2018} at both near- and far-field planes.

\par Upon propagation through free space via a Fresnel kernel to a distance $z$, the wavefunction retains its double Gaussian structure, with the correlation widths evolving as $\sigma_\pm(z) = \sqrt{\sigma_\pm^2 + \frac{4\pi^2 z^2}{\lambda^2}}$ \cite{chan_transverse_2007,just_transverse_2013,Tasca2009}, where $\lambda$ is the wavelength of the down-converted photons. Notably, there exists a special propagation plane at a distance $z_p = \pm \frac{2\pi}{\lambda} \sigma_+\sigma_-$ \cite{Tasca2009}, at which all spatial amplitude correlations vanish, and the entanglement manifests purely through phase correlations \cite{chan_transverse_2007}. The quantum state at this plane is given by:

\begin{widetext}
\begin{eqnarray}
\label{eqn:DG phase}
\psi_p(x_1,x_2)  & = \mathcal{N} \exp{\left(-\frac{1}{4}\left(\frac{x_1^2+x_2^2}{\sigma_+^2+\sigma_-^2}\right)\right)}\exp{\left(\frac{i}{4\left(\sigma_+^2+\sigma_-^2\right)}\left((x_1-x_2)^2\frac{\sigma_+}{\sigma_-} + (x_1+x_2)^2\frac{\sigma_-}{\sigma_+}\right)\right)} \nonumber\\
&=\mathcal{N} \exp{\left(-\frac{1}{4}\left(\frac{x_1^2+x_2^2}{\sigma_+^2+\sigma_-^2}\right)\right)} \exp{\left(\frac{i\pi}{2\lambda z_p}(x_1^2+x_2^2)\right)}\exp{\left(-\frac{i x_1x_2}{2\sigma_+\sigma_-}\left(\frac{\sigma_+^2-\sigma_-^2}{\sigma_+^2+\sigma_-^2}\right)\right)}
\end{eqnarray}
\end{widetext}

\par Such states are experimentally accessible, and their realization can be verified in multiple ways. One method is to monitor the loss of amplitude correlations during propagation, for instance through the \textit{Fedorov ratio} \cite{Fedorov_2005,fedorov_gaussian_2009}, which becomes unity at this plane \cite{Tasca2009,reichert_quality_2017,Rounak2025}. Alternatively, direct methods such as holographic phase reconstruction \cite{abouraddy_quantum_2001,Zia2023} or interference-based techniques \cite{Rounak2025} can be employed to characterize the state. A key insight from the above expression (Eq.\ref{eqn:DG phase}) is revealed by rewriting certain coefficients as $A = \frac{1}{4(\sigma_+^2 + \sigma_-^2)}$ and $B = \frac{1}{2\sigma_+\sigma_-} \cdot \frac{\sigma_+^2 - \sigma_-^2}{\sigma_+^2 + \sigma_-^2}$. This allows us to express the quantum state as:

\begin{equation}
\label{eqn:DG phase rewritten}
\psi_p(x_1,x_2) = \mathcal{N} \exp{\left(\frac{ik}{4z_p}\left(x_1^2+x_2^2\right)\right)}\psi_{p^2}(x_1,x_2)
\end{equation}

\par Thus, from the experimentally accessible phase-entangled state, one can obtain the pure phase-entangled state by removing the uncorrelated quadratic phase term of each photon in the wavefunction. Fortunately, this can be achieved using single-lens imaging, as will be explained in the next subsection(sec.\ref{subsec:state prep}). Once the state is prepared, its unique correlation structure can be verified through experimental techniques, which will be discussed in \ref{subsec:Measurement theory}.
\subsection{State preparation}
\label{subsec:state prep}
To streamline the computation, we introduce quantum operators that describe spatial quadratic phase fronts and scaling in the position basis:

\begin{eqnarray}
\label{eqn:Ops}
&\hat{Q}[c,k]_x f(x)\equiv \exp{\left(-\frac{i\pi cx^2}{\lambda}\right)}f(x) \nonumber \\
&\hat{\mathcal{V}}[s]_x f(x)\equiv \sqrt{s}~f(sx)
\end{eqnarray}

\par Here, $s$ is a dimensionless scaling factor, and $c$ has units of inverse length. Using these operators, Eq.\ref{eqn:DG phase rewritten} can be compactly expressed as:

\begin{equation*}
\label{eqn:compressed phase}
\psi_{p}(x_1,x_2) =  \hat{Q}\left[\frac{1}{2z_p},k\right]_{x_1}\hat{Q}\left[\frac{1}{2z_p},k\right]_{x_2}\psi_{p^2}(x_1,x_2)
\end{equation*}

\par In optical theory, it is well known that a thin lens of focal length $f$, used to image an electromagnetic field located at object distance $u$, produces a scaled image at distance $v = \frac{uf}{u - f}$ with an additional quadratic phase factor dependent on $u$ and $f$. The equivalent quantum operator for such single-lens imaging in the position basis is:
\begin{equation}
\hat{\mathcal{I}}[u,f,k]_x \equiv \hat{\mathcal{V}}\left[1-\frac{u}{f}\right]_x\hat{Q}\left[\frac{1}{u-f},k\right]_x
\end{equation}
This formulation clearly shows that real imaging $(u > f)$ introduces a positive quadratic phase, while virtual imaging $(u < f)$ imparts a negative one. Therefore, by applying a “single-lens virtual imaging” operator with a specific object distance to the phase-entangled state in Eq.\ref{eqn:compressed phase}, we can exactly cancel the positive quadratic phase term and recover the pure phase-entangled state in a virtual image plane. The required object distance for this cancellation is simply $u = f - 2z_p$.

\par To retrieve this virtual state on a real plane, we employ a $4F$ imaging system with magnification $M$. This system can be modelled using the scaling operator (defined in Eq.\ref{eqn:Ops}) with $s = -\frac{1}{M}$. Thus, the full sequence of operations required to transform $\psi_p$ into a scaled version of $\psi_{p^2}$—which we refer to as the pure phase plane state, denoted by $|P^3\rangle$ is:

\begin{widetext}
\begin{eqnarray}
\label{eqn:operator}
&\psi_{P^3}(x_1,x_2)= \hat{\mathcal{V}}\left[-\frac{1}{M}\right]_{x_1}\hat{\mathcal{V}}\left[-\frac{1}{M}\right]_{x_2}\hat{\mathcal{I}}\left[f - 2z_p, f, k\right]_{x_1}\hat{\mathcal{I}}\left[f - 2z_p, f, k\right]_{x_2} \psi_{p}(x_1,x_2) \nonumber \\
&\quad = \hat{\mathcal{V}}\left[-\frac{2z_p}{Mf}\right]_{x_1}\hat{\mathcal{V}}\left[-\frac{2z_p}{Mf}\right]_{x_2} \psi_{p^2}(x_1,x_2)
\end{eqnarray}
\end{widetext}

\begin{figure}[!htbp]
    \centering
    \includegraphics[width=0.8\linewidth, trim = 0cm 0cm 0cm 0cm, clip]{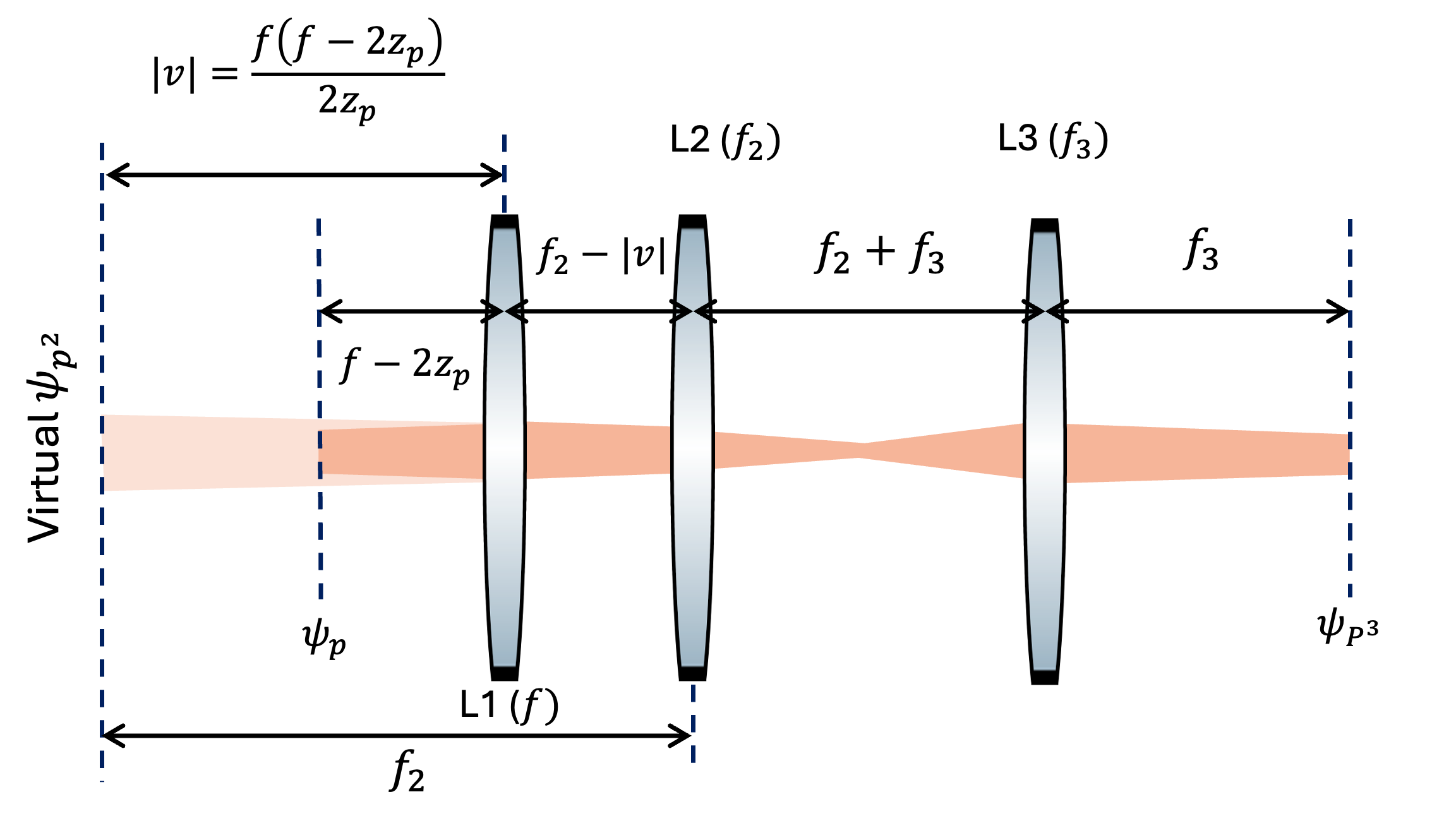}
    \caption{Schematic for preparation and measurement of $|P^3\rangle$ (Eq.\ref{eqn:operator}) state. First lens$L1(f)$ creates virtual image with object distance $u = f-2z_p$, followed by $4F$ imaging constituted by $L2(f_2)\text{ and }L3(f_3)$ with  magnification $\frac{f_3}{f_2}$. Thus effective magnification $= \frac{ff_3}{2z_pf_2}$}
    \label{fig:state schematic}
\end{figure}

This entire sequence of operations is illustrated schematically in Fig.~~\ref{fig:state schematic}.

 The virtual plane obtained after first lens at imaging distance $v = -\frac{(f-2z_p)f}{2z_p}$. Hence,  if the $4F$ system has  lenses of focal length $f_2$ and $f_3$ giving rise to $M = \frac{f_3}{f_2}$, then the $4F$ system must be placed at a distance of  $f_2-|v|$ from the virtual imaging lens, so that the virtually imaged plane remains at a distance  $f_2$ from the $4F$ system. The added constraint for successful generation is that $f_2>|v|$ to obtain a real image. The density measurement at this location should give the Fedorov ratio $=1$\cite{fedorov_gaussian_2009} since it's essentially a phase-entangled state. Its easy to observe that scaling only changes the effective value of parameters $A$ and $B$.
\subsection{Theoretical schematic of partial Fourier transform measurement}
\label{subsec:Measurement theory}
For a comprehensive proof of the state, we execute the \textit{PFT} of the state by a split measurement scheme, where the biphoton wavefunction is split in two paths using a beamsplitter. Only those events are considered where one photon passes on each path\cite{zhang_background_2024}. 
\begin{figure}
    \centering
    \includegraphics[width=0.5\linewidth,trim = 1.5cm 4.5cm 0.9cm 2cm,clip]{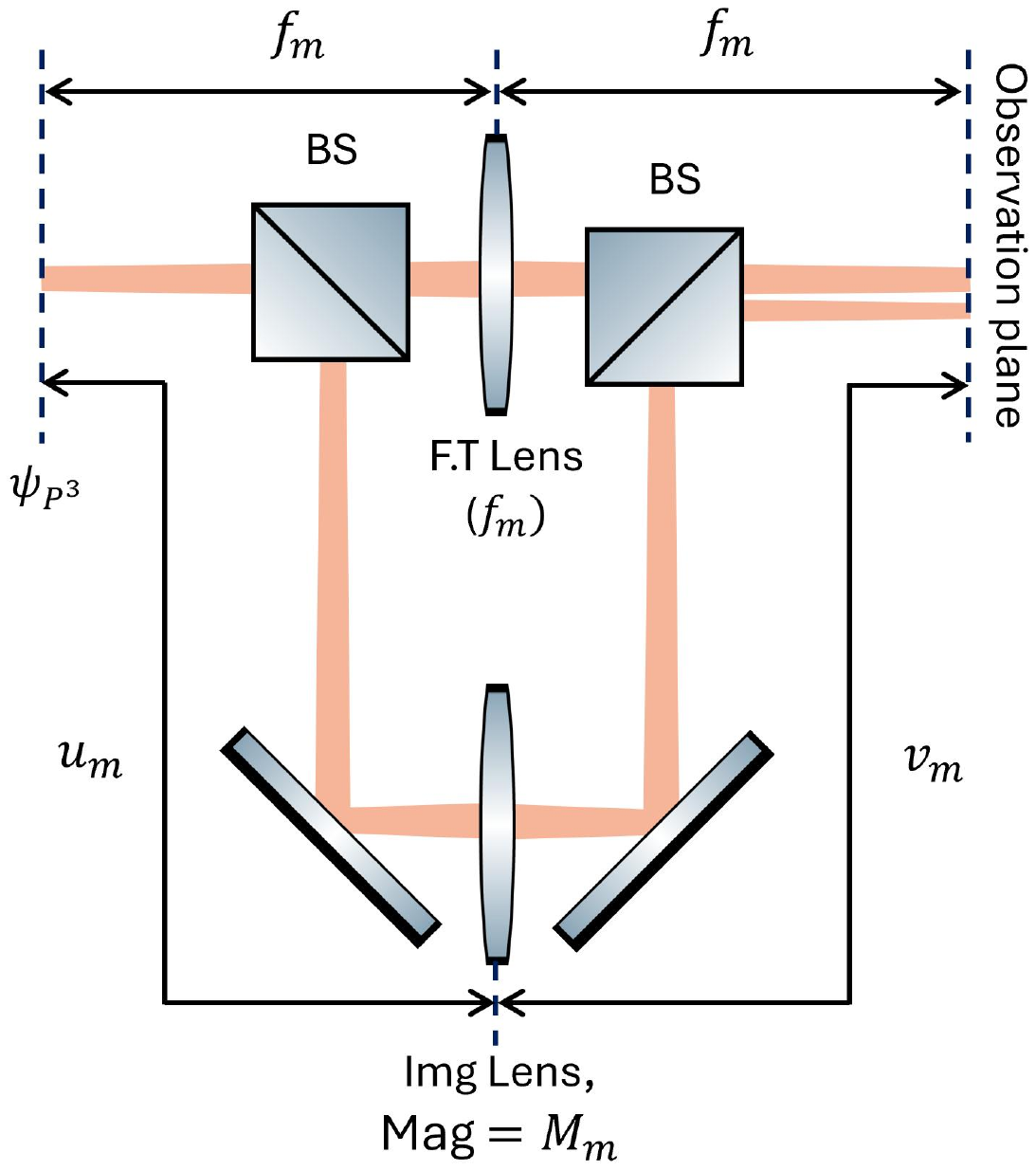}
    \caption{The schematic setup for  partial Fourier transform with F.T. Lens $(f_m)$ performing  $2f$ optical Fourier transform in the transmitted path while Img lens performing single-lens imaging with magnification $M_m$  in the reflected path. }
    \label{fig:measurement schematic}
\end{figure}
\par As shown in Fig.~~\ref{fig:measurement schematic}, one path consisting of a single  lens of focal length $f_m$ in a $2f$ configuration performs a Fourier Transform, while the other path performs single lens imaging with magnification $M_m$.  Mathematically, the density function $(\rho_m)$ produced at the observation plane in Fig.~~\ref{fig:measurement schematic} in terms of spatial variables mapping position $(x_p)$ and momentum $(x_k)$  \cite{Goodman1996} is given as:
\begin{eqnarray}
\label{eqn:PFT density}
    & \rho_m(x_k,x_p) = \nonumber\\ & \Bigl| \frac{1}{\sqrt{\lambda f _mM_m}}\int{\psi_{P^3}(x'_k,\frac{x_p}{M_m}) \exp{\left(-2\pi i\frac{x'_kx_k}{\lambda f_m}\right)}~dx_k'}\Bigr|^2
\end{eqnarray}

Using Eq.\ref{eqn:PFT}, the explicit form can be written as:
\begin{eqnarray}
\label{eqn:Practical PFT}
    &\rho_m(x_k,x_p) = \nonumber \\&\frac{2}{\lambda f_m M_m}~\exp{\left(-2 A'\frac{x_p^2}{M_m^2}\right)}\exp{\left(-\frac{\left(\frac{2\pi x_k}{\lambda f_m}+B'\frac{x_p}{M_m}\right)^2}{2A'}\right)}
\end{eqnarray}
where $A' = \frac{4z_p^2}{M^2f^2}A\text{ and }B' = \frac{4z_p^2}{M^2f^2}B$.
\begin{figure}
    \centering
    \includegraphics[width=0.8\linewidth,trim = 0.25cm 0.25cm 0.25cm 0.25cm,clip]{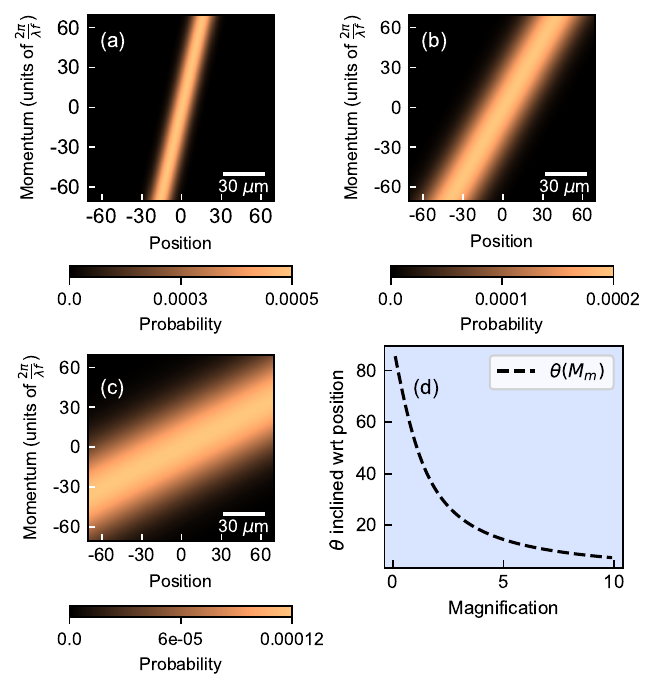}
    \caption{Theoretical estimates of joint position of  one -momentum of other measurement. The two photon densities (plot of Eq.\ref{eqn:Practical PFT}) for magnification $M_m=~$(a) $-0.3$ (b) $-0.75$ (c) $-2.5$. (d) The variation of tilt of the density function with respect to position axis $\theta$ as a function of  magnification $M_m$.}
    \label{fig:Theory Plots}
\end{figure}
 \par Typical forms of this density are shown in Fig.~~\ref{fig:Theory Plots}. For a fixed value of $A', B' \text{ and } f_m$, the functional form is a generalised two-dimensional Gaussian \cite{2D_gaussian}, whose widths and  orientation with respect to the position axis are a function of $M_m$. This is depicted in Fig.~~\ref{fig:Theory Plots}(a),(b)  and (c). The major tractable here is the tilt angle$(\theta)$ with respect to the position axis of the distribution, whose exact analytical form is given by:
\begin{eqnarray}
\label{eqn:theta func}
    &\theta = \frac{1}{2}\tan^{-1}\left(\frac{2b}{a-c}\right)\text{, where}\nonumber\\
    &a = \frac{2\pi^2}{\lambda^2f_m^2A'}\text{, }b=\frac{\pi B'}{\lambda f_mM_mA'}\text{, }c = \frac{2A'}{M_m^2}+\frac{B'^2}{2M_m^2A'}
\end{eqnarray}
In our experimental verification, we'll use this fact to establish the inherent nature of position-momentum correlation and argue that it's not an artefact of preparation or measurement scheme. The corresponding spread in position and momentum can be found by eigenvalue analysis of matrix $\big(\begin{smallmatrix}
  a & b\\
  b & c
\end{smallmatrix}\big)$\cite{2D_gaussian}. The variation in tilt angle is shown in Fig.~~\ref{fig:Theory Plots}(d).

\section{Experiments and Results}
\label{sec:Exp descriptor}
Experimentally, initial calibration of the quantum state is important, since information about the position correlation width$(\sigma_-)$ and momentum anti-correlation width$(\sigma_+)$ is enough to tell us the steps to construct the pure phase entangled  state. We achieve this by following a procedure similar to \cite{Edgar2012}. Once estimated, we prepare the state and do the exact kind of  measurement as  described in \ref{sec:theory}. 
\subsection{Estimation of parameters and initial tests}
\label{sec: Exp params}
\begin{figure}[!htbp]
    \centering
    \includegraphics[width=0.9\linewidth,trim = 1.5cm 0.1cm 1.5cm 1cm,clip]{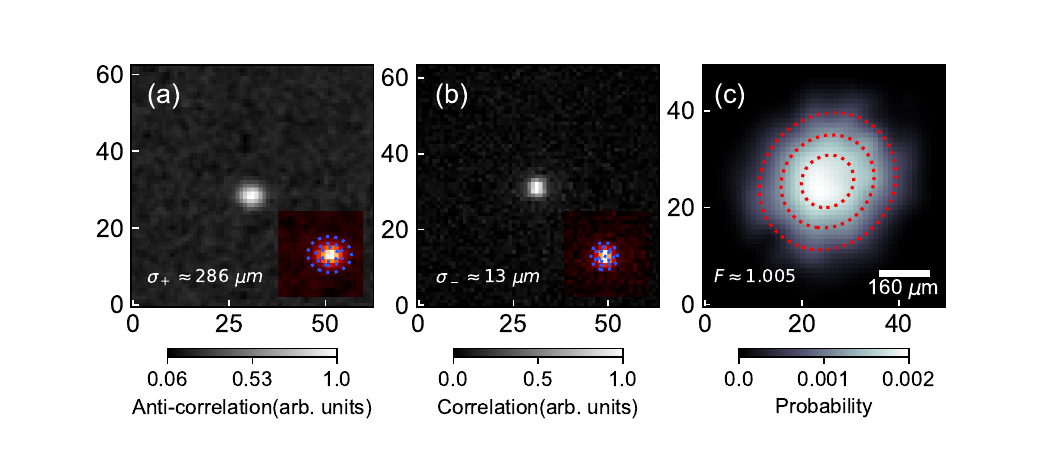}
    \caption{Experimental estimation of  initial parameters  $\sigma_\pm$  for  calibrating initial DG wavefunction(Eq.\ref{eqn:DG}) and and estimating the state after three lens system shown in Fig.~~\ref{fig:state schematic}. (a) Momentum anti-correlation image providing $\sigma_+\approx 286~\mu$m). (b)Position correlation width($\sigma_-\approx13~\mu$m) . (c) Background corrected density\cite{Rounak2025} of pure phase entangled wavefunction with estimated $F\approx1.005$.}
    \label{fig:initial measurement}
\end{figure}

The key parameters of this experiment are the practical values of $\sigma_{\pm}$ that determine the initial DG state (Eq.\ref{eqn:DG}). This is obtained by estimating the position correlation width in near-field giving  $\sigma_-$ and momentum anti-correlation width in far-field giving $\sigma_+$\cite{schneeloch_introduction_2016}. Experimentally, correlation width is obtained by imaging the crystal center, acquiring multiple photon-pair frames, and autocorrelating these frames  to obtain a Gaussian with width  $\sigma_-$. While the anti-correlation width is obtained  by observing the crystal center with a $2F$ Fourier transforming setup, acquiring multiple frames and autoconvolving them to obtain a Gaussian with width $\sigma_+$.\cite{Edgar2012}. The end process of these is shown in Fig.~~~\ref{fig:initial measurement}(a) and  (b) respectively, where the Gaussian fits give $\sigma_-\approx13~\mu\text{m and }\sigma_+\approx286~\mu \text{m}$. Using these two parameters, we compute the location of the phase plane $z_p \approx 2.97$ cm and approximate values $A\approx3\times10^{-6} ~\mu\text{m}^{-2}$,  $B\approx128\times 10^{-6}~\mu\text{m}^{-2}$. To justify that only phase correlations exist at the $|P^3\rangle$ plane, we first experimentally construct the setup as per Fig\ref{fig:setup_main} and measure the lack of amplitude correlations in this plane. We compute the Fedorov ratio $(F)$ at this plane, which is the ratio of  the photon marginal distribution$(\sigma_{x_1})$ to the spread of one photon's conditional distribution $(\sigma_{x_1|x_2})$, that is $F= \frac{\sigma_{x_1}}{\sigma_{x_1|x_2}}$\cite{fedorov_gaussian_2009}. Ideally, we should obtain  $F = 1$ for a spatially uncorrelated state\cite{chan_transverse_2007}, which is experimentally constructed from the two-photon density function as shown in Fig.~~\ref{fig:initial measurement}(c). The density is constructed  by acquiring $600,000$ frames at $|P^3\rangle$ plane using an Electron Multiplying Charge Coupled Device (EMCCD) and resolving the photon counts\cite{Rounak2024}, followed by statistical construction of density using the procedure described in\cite{defienne_general_2018}. The density is fit with a DG function  since the spatial amplitude is still  a double Gaussian and the experimental Fedorov ratio is $F \approx 1.005$. Thus, the photon pairs are uncorrelated in amplitude, and all of the correlation remains  in the phase. 

\subsection{Experimental setup}
\label{sec:Actual exp}
\begin{figure*}[!htbp]
    \centering
    \includegraphics[width=0.9\linewidth,trim = 0.75cm 3.5cm 0.75cm 3.3cm ,clip]{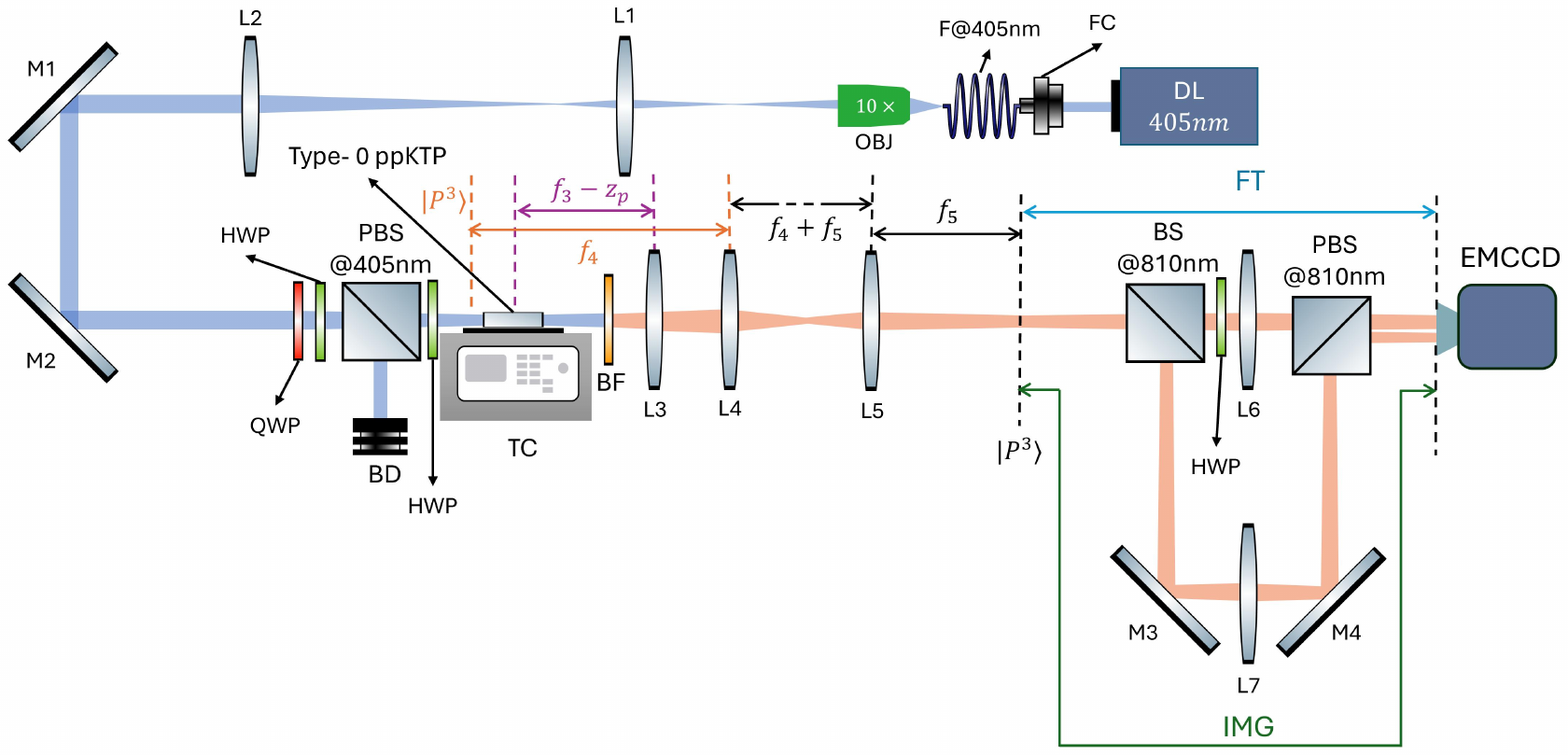}
    \caption{Full schematic setup for the preparation and study of pure phase entanglement. DL: Diode Laser ($\lambda_p = 405\pm0.1$ nm), FC: in-coupling fiber  collimator, F@$405$nm: Single mode FC/PC fiber, OBJ: $10\times$ objective, WD $\approx 3.7$ mm, L1 ($f_1=2.5$ cm), L2 ($f_2=20$ cm), M1, M2: relaying lenses and mirrors that bring beam waist($\approx 450~\mu$m) on crystal center. HWP, QWP, PBS@$405$nm: Zero Order ,$405$ nm centred half-wave plate, quarter-wave plate and polarising beam splitter for vertically polarising pump beam, TC: Temperature Controller maintaining $28.6\degree $C, BF: Two bandpass filters centred at $810$ nm with bandwidths $3$ nm and $10$ nm respectively. L3 $(f_3 = 10\text{ cm})$ ,L4 $(f_4 = 15\text{ cm})$, L5 $(f_5 = 12.5\text{ cm})$: State preparation lenses placed as per schematic in Fig.~~\ref{fig:state schematic} , BS, HWP, PBS, M3, M4, ,L6$(f_m = 15\text{ cm})$, L7$(M_m)$, : Measurement schematic as per Fig.~\ref{fig:measurement schematic} . BS is a beam splitter. Lenses L7 and L8 do imaging with magnification $M_m$ and a Fourier transform, with focal length $f_m$ respectively. HWP: Zero-order $810$ nm half-wave plate and PBS: Polarising beam splitter ensure that all photons reach the observation plane without loss, EMCCD: Electron multiplying charged coupled device.}
    \label{fig:setup_main}
\end{figure*}
 The schematic of the practical setup is shown in Fig.~ \ref{fig:setup_main}. A Gaussian pump beam $(\text{Wavelength = }405 \text{nm, vertically polarized})$ is achieved by using a single-mode fibre. The beam waist is relayed to  the crystal center (type-0 ppKTP) after suitable polarization optics. The location and size of the beam waist are estimated by tracking the beam's propagation  \cite{Vikas2024}, and the estimated waist $\approx 450\mu$m. The smallest beam waist location corresponds to a zero phase Gaussian field, required for the generation of initial DG state \cite{chan_transverse_2007}.  The three lenses are placed to create the pure phase entangled state at plane $|P^3\rangle$ following the schematic in Fig.~\ref{fig:state schematic}. The two measurement paths are constructed using schematic of Fig.~\ref{fig:measurement schematic} and acquired concurrently on the EMCCD. By ensuring that frames are receiving photon pairs with a mean rate $\lessapprox 0.15$ per pixel per frame, we can be assured that for an observed pair, one photon is measured in position basis while other in momentum. The approximate magnification value was estimated by placing a USAF target at the $|P^3\rangle$ plane and observing it on EMCCD through the imaging path.  

\par The position to momentum correlation curves are constructed  by summing the data over the $y-$axis and then computing the two-dimensional correlation from the $x-$axis counts $(c_{x_i},i=\{k,p\})$:
\begin{eqnarray}
\label{eqn:PFT density exp}
    &\rho_{m}(x_k,x_p) = \langle c_{x_k}c_{x_p}\rangle - \langle c_{x_k}\rangle \langle c_{x_p}\rangle \nonumber \\ [4pt]
    &\langle c_{x_k}c_{x_p}\rangle = \frac{1}{N}\sum_{j=1}^N{c^{(j)}_{x_k}c^{(j)}_{x_p}}\nonumber \\ [4pt]
    &\langle c_{x_k}\rangle \langle c_{x_p}\rangle \approx \frac{1}{N-1}\sum_{j=1}^{N-1}{c^{(j)}_{x_k}c^{(j+1)}_{x_p}}\nonumber\\
\end{eqnarray}
Where $c^{(j)}_{x_i},i = \{k,p\}$ is photon count at position $x_i \text{ for } j^{\text{th}}$ frame number. The density is self-normalised to unity. The form of $\langle c_{x_k}\rangle \langle c_{x_p}\rangle $ used in Eq.\ref{eqn:PFT density exp} is computationally light to execute and works because of very short temporal correlation of photon pairs along with their very low mean rate\cite{Hugo2019} . 
\subsection{Results and discussion}
\label{subsec:results}
\begin{figure}
    \centering
    \includegraphics[width=0.95\linewidth,trim = 1.5cm 1cm 8cm 1cm,clip]{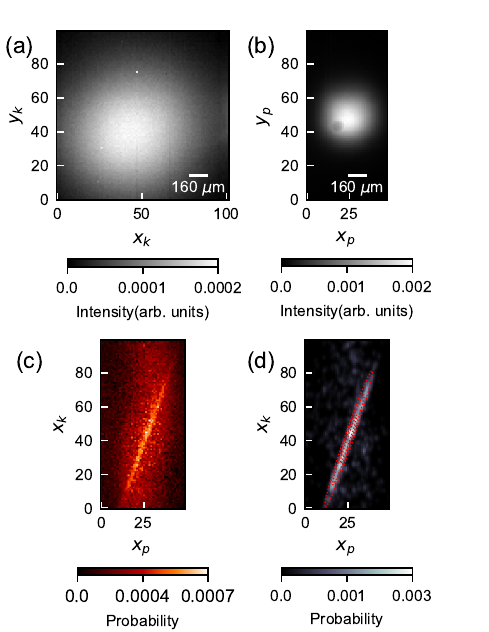}
    \caption{Experimental estimation of  $\rho_m(x_k,x_p)$  for magnification $M_m \approx -0.5$. (a) Measured mean beam for arm measuring momentum on one photon in space variables$(x_k)$. (b) Measured  mean beam for arm measuring position of other photon with magnification $(x_p)$. (c) Computed density $\rho_m(x_k,x_p)$ from $500,000$ frames. (d) Processed $\rho_m$ after statistical cleaning. The dotted contour is  a 2D Gaussian fitted with computed inclination $\theta \approx 70.9\degree$
    All probability images are self normalized to unity.}
    \label{fig:Mag05}
\end{figure}

The acquired one photon marginal beams averaged over $500,000$ frames are shown in Fig.~\ref{fig:Mag05}(a) and (b), featuring momentum measurement with the lens of  focal length $f_m = 15$ cm $(x_k,y_k)$ and position measurement with magnification $M_m\approx -0.5~(x_p,y_p)$, respectively. The computed experimental density function from these frames is shown in Fig.~\ref{fig:Mag05} (c). Due to finite sampling of the wave function, one can see fluctuation noise as well as a low-frequency background that is functionally similar to the marginal distribution of the photons. Statistical noise cleaning techniques described in Appendix \ref{app:statistical cleaning} leading to Fig.~\ref{fig:Mag05}(d). We fit this  distribution with a generalised two-dimensional Gaussian\cite{2D_gaussian} as shown by red dotted contour overlay in Fig.~\ref{fig:Mag05}(d). The visible strong correlation endorses the the existence of a pure phase entangled state. The fact that the JPD has an inclination$>45\degree$ is consistent with theoretical predictions for $M_m \approx 0.5$.  

\par
\begin{figure}[!htbp]
    \centering
    \includegraphics[width=1\linewidth,trim = 1.5cm 0.25cm 2cm 0.9cm,clip]{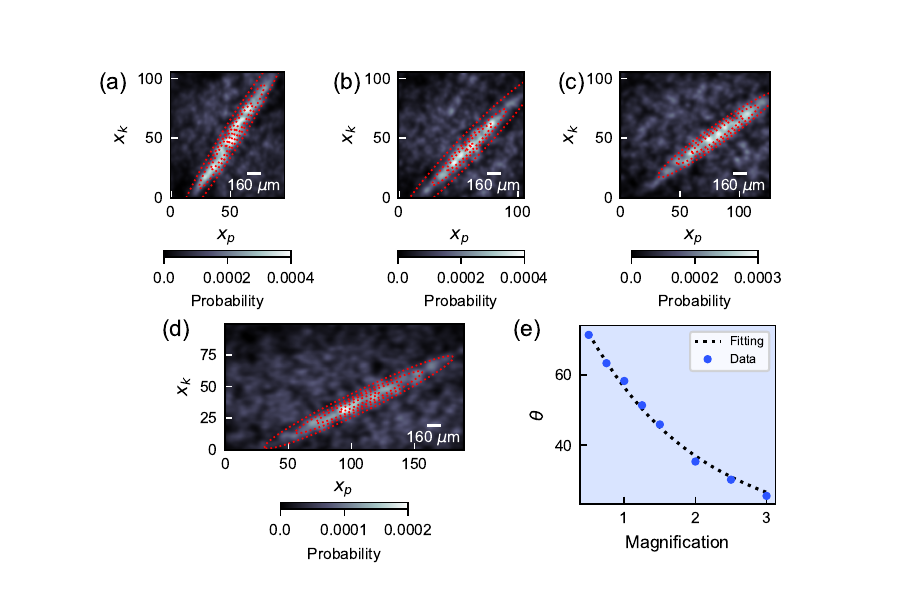}
    \caption{Variation of  correlation axis tilt $(\theta)$ with respect to position axis as a function of changing magnification $M_m$ (estimated using USAF target at $|P^3\rangle)$. The four instances shown here are the $\rho_m(x_k,x_p)$ densities for various $M_m \approx$ (a) $1.0$ (b) $1.5$ (c) $2.0$ (d) $2.5$. These are fitted with a 2D Gaussian shown by red dotted contour overlays. (e) Variation of experimental inclination (blue dots) with respect to magnification. Data is fitted with Eq.\ref{eqn:theta func} with net three-lens magnification $(M)$ as fitting parameter. Theoretically estimated $M\approx1.39$, experimentally fitted $M\approx 1.29$.}
    \label{fig:Mag variation}
\end{figure}

 A few further magnifications were utilized to verify this correlation as shown by the densities in Fig.~\ref{fig:Mag variation}(a)-(d). The experimental variation of inclination as a function of $M_m$ shown in Fig.~\ref{fig:Mag variation}(e) is in excellent agreement with the theoretical prediction Eq.\ref{eqn:theta func} (dotted black line). The fit parameter $M$ (occurring in the expression of $A'\text{ and }B'$)  $7\%$ of the theoretically estimated value. 
\par The above physical results can be mathematically formalized by studying the conditioned transverse momentum of one photon. The marginal feature is obtained from the momentum representation of the wavefunction in Eq.\ref{eqn:PFT} by computing $\rho(p_1)=\int{|\phi_{p^2}(p_1,p_2)|^2~dp_2}$, which is a zero-centered Gaussian with a width $\sigma_{p_1} = \hbar \sqrt{A+B^2/4A}$. For measuring the conditional momentum mean and width, we first express the pure phase-entangled quantum state in the basis of the position of one photon and the momentum of the other:
\begin{equation}
|\psi_{p^2}\rangle = \int dp_1dx_2~\psi_{p^2}(p_1,x_2)~|p_1,x_2\rangle
\end{equation}
where $\psi_{p^2}(p_1,x_2)$ is given by Eq.\ref{eqn:PFT}. From the theory of “projection-valued measurements,” it follows that if the outcome of a position measurement is $x_2 = \text{x}$, the corresponding projection operator is $\hat{P}{\text{x}} = |\text{x}\rangle\langle \text{x}|$. Applying $\hat{P}{\text{x}}$ to the state $|\psi_{p^2}\rangle$ and normalizing yields:
\begin{eqnarray}
&|\psi_{p^2}\rangle_{\text{x}} = \frac{\hat{P}_{\text{x}}|\psi_{p^2}\rangle}{\sqrt{\langle\psi_{p^2}|\hat{P}_{\text{x}}^{\dagger}\hat{P}_{\text{x}}|\psi_{p^2}\rangle}}\nonumber \\
&\Rightarrow \frac{\int{dp_1~\psi_{p^2}(p_1,\text{x})~|p_1,\text{x}\rangle}}{\sqrt{\int{dp_1~|\psi_{p^2}(p_1,\text{x})|^2}}}\nonumber \\
& \propto \int{dp_1~\exp{\left(-\frac{(p_1+\hbar B\text{x})^2}{4A\hbar^2}\right)~|p_1,\text{x}\rangle}}
\end{eqnarray}

\par conditional momentum density $\rho(p_1|\text{x}) = |\langle p_1,\text{x}|\psi_{p^2}\rangle_{\text{x}}|^2$, is a Gaussian with mean $\langle p_1|\text{x}\rangle = -\hbar B\text{x}$ and standard deviation $\sigma_{p_1|\text{x}} = \hbar\sqrt{A}$.

\par One can observe that $\sigma_{p_1}>\sigma_{p_1|\text{x}}$, hence marginally, each photon possesses the maximum transverse momentum spread permitted by the source of the photon pairs and is therefore only partially coherent. In contrast, the conditional spread is smaller, indicating that the post-measured photon exhibits high transverse momentum coherence. The marginal coherence declines as the state becomes more and more entangled. This can be practically understood understood by examining the forms of $A$ and $B$ in terms of $\sigma_\pm$, when the state is generated from a phase-entangled source, in the asymptotic limit of a high Schmidt number (practically realized for $\sigma_+\gg \sigma_-$). It can be shown under this limit $\sigma_{p_1}\approx 1/2\sigma_-$ and $\sigma_{p_1|\text{x}}\approx 1/2\sigma_+$. hence $\sigma_+\gg\sigma_-$ implies $\sigma_{p_1|\text{x}} \ll \sigma_{p_1}$.

\par
\begin{figure}[!htbp]
    \centering
    \includegraphics[width=1\linewidth]{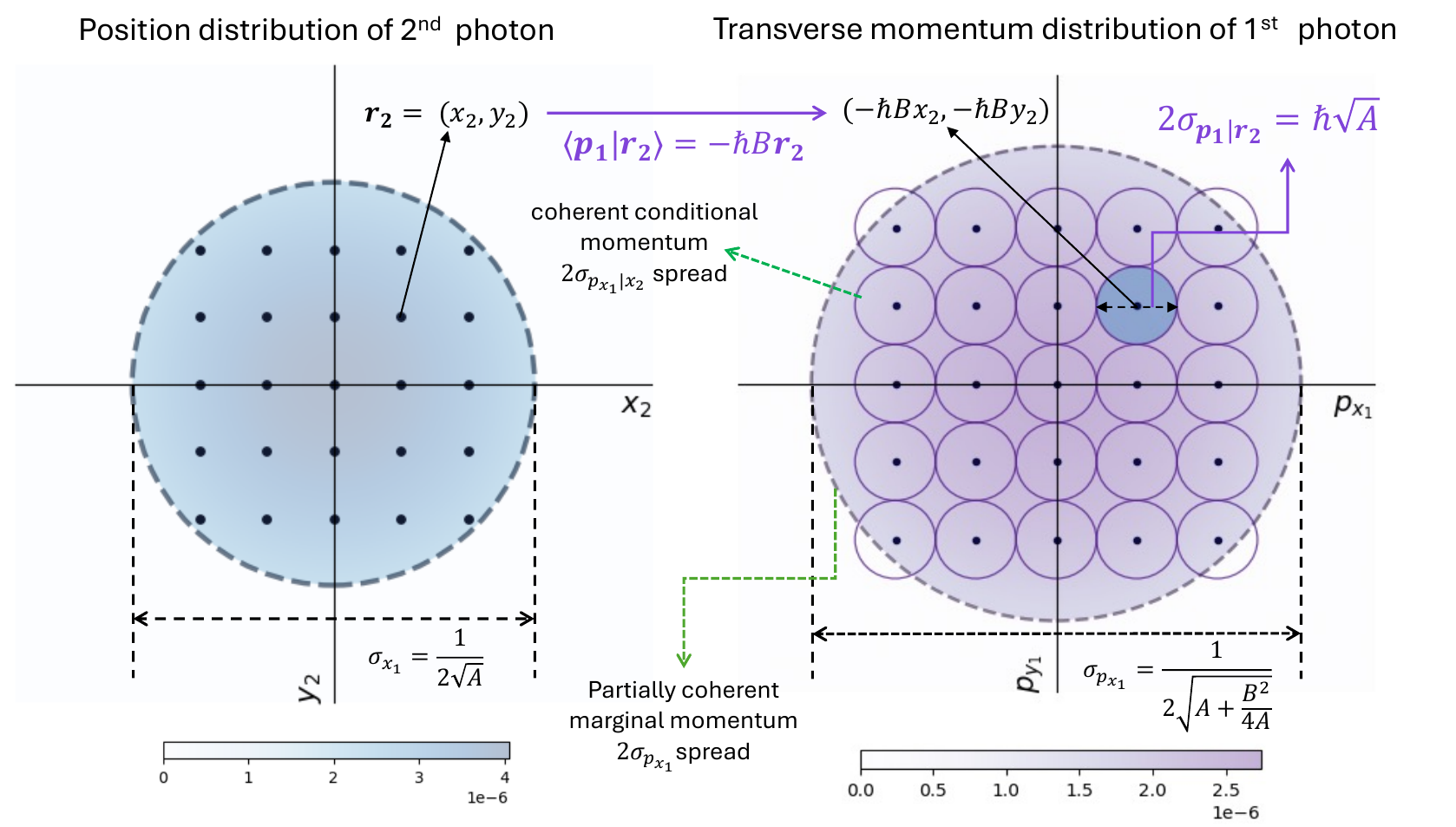}
    \caption{Pure phase entangled state showing increase in one photon transverse momentum coherence on measurement of position of the other.}
    \label{fig:physics}
\end{figure}

 In a two-dimensional scenario, this situation can be visualized as a \textit{broadband transverse momentum} one-photon Gaussian beam composed of multiple \textit{narrowband sources} emitting randomly. A position measurement on one photon effectively enhances the spatial coherence of the other. This duality between coherence in entangled systems and partially coherent light has been previously studied \cite{saleh_duality_2000}. A pictorial representation of this concept in two dimensions is illustrated in Fig.~\ref{fig:physics}. The most natural way of quantitatively thinking about this is in a fashion similar to birth zones \cite{schneeloch_introduction_2016,Vikas2025}. Birth zones are defined as independent correlated photon generation regions embedded in a Gaussian region defined by the spread of the pump. It quantifies the 2-photon coherence of SPDC process. For a DG state (Eq.\ref{eqn:DG}), one can define a ``\textit{birth zone number}" $N = \frac{\text{Spread of  pump}}{\text{Pair correlation width}} =\frac{\sigma_+}{\sigma_-}$. More naturally, it defines the maximum number of independent two-photon Gaussian regions that can defined in the area of the marginal beam\cite{Vikas2025}.  In the asymptotic limit for high Schmidt number pure phase entangled system, $\frac{2\sigma_{p_1}}{2\sigma_{p_1|\text{x}}} \approx \frac{\sigma_+}{\sigma_-} = N$. 
\section{Conclusion}
\label{sec:conclusion}
In conclusion we have established a practical scheme for the pure phase-entangled state previously proposed only in theoretical frameworks. We have shown that the wavefunction can be interpreted as a quantum system that reconfigures its pairwise correlation into a single-photon wave, whose central transverse momentum and bandwidth are governed by conditional position measurement of other. Furthermore, we demonstrate that by tuning position measurement parameters, we gain control over the span and structure of the other's coherence properties, enabling direct applicability across various quantum technologies.

\par The pure phase entangled state could have variety of applications. the one immediate choice could be heralded imaging modalities \cite{Morris2015,Aspden2016}, where the position information of one photon can be used to infer the momentum characteristics of its twin, which interacts solely with the sample. This could  become particularly useful in probing disordered media, where knowledge of the source’s initial statistics is crucial \cite{Gnatiessoro2019,Bajar2025} or in Fourier ptychography \cite{Zheng2013,Zheng2021,Konda2020}, whose quantum analogue has been demonstrated in \cite{Aidukas2019} and a similar method can be proposed for this system. Beyond its position-momentum correlation, the wave-function's highly correlated phase front offers opportunities in quantum imaging, particularly within phase-contrast microscopy paradigms. To enable this, the initial wavefront distribution must be carefully configured—an objective achievable through quantum holography \cite{abouraddy_quantum_2001}, which provides direct access to the system’s quantum phase distribution \cite{Hugo2019,Zia2023}. 
\section{Acknowledgement}
We acknowledge the Department of Atomic Energy, Government of India, for funding for Project Identification No. RTI4002 under DAE OM No. 1303/1/2020/R\&D-II/DAE/5567, Ministry of Science and Technology, India. We also acknowledge Saptak Mandal for his support in realising this project.
\appendix
\section{Statistical Cleaning of Data}
\label{app:statistical cleaning}
Due to finite sampling, the marginal distribution of photons turns out as noise of  low-frequency embedded in statistical noise. One method to remove this noise is to use the concept of excess two-photon correlation \cite{Saleh2000} which extracts the pure two-photon contribution $\Delta G^{(2)}$ defined as:
\begin{eqnarray*}
    &\Delta G^{(2)}(x_k,x_p) = \rho_m(x_k,x_p) \nonumber\\ &- \int{\rho_m(x_k,x_p)~dx_p} ~\otimes\int{\rho_m(x_k,x_p)~dx_k} 
\end{eqnarray*}
\par
\begin{figure}
    \centering
    \includegraphics[width=1\linewidth,trim = 1cm 1cm 1cm 1cm,clip]{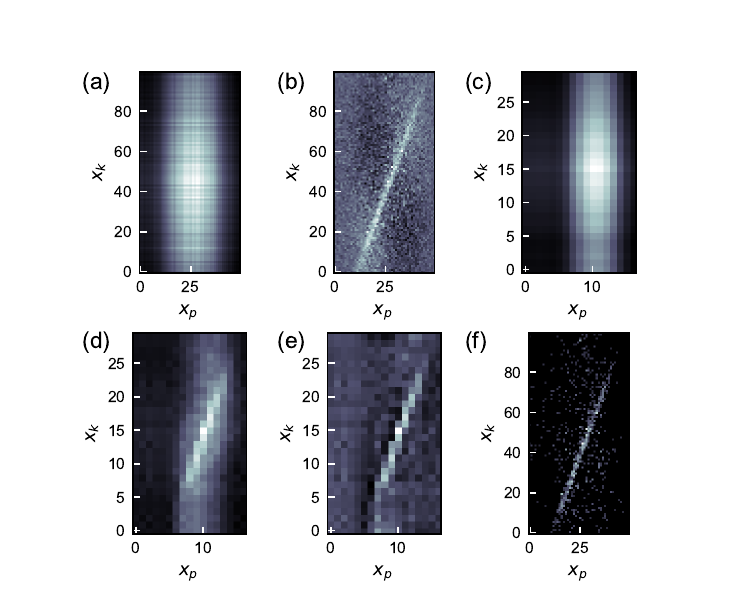}  
    \caption{Various images showcase the cleaning procedure for density $\rho_m$. (a) Marginal image $\rho_M(x_k,x_p) = \sum_{x_k}\rho_m(x_k,x_p) \otimes \sum_{x_p}\rho_m(x_k,x_p)$. (b) Computed  excess two-photon correlation $\Delta G^{(2)}$. Daubechies wavelet decomposed approximate image of  (c) $\rho_M$ (d) $\rho_m$. (e) Fourier filtered approximate  $\rho_m$ with filter bandwidth decided by power spectral analysis of approximate $\rho_M$. (f) Reconstructed $\rho_m$ using wavelet reconstruction using filtered approximate $\rho_m$.   }
    \label{fig:stat cleaning}
\end{figure}

 Here, the first term is the two photon density. The second term represents the \textit{marginal image} in the $(x_k,x_p)$ space which is computationally calculated as $\rho_{M}(x_k,x_p) = \sum_{x_k}\rho_m(x_k,x_p) \otimes \sum_{x_p}\rho_m(x_k,x_p)$.  As shown in Fig.~\ref{fig:stat cleaning}(a), $\rho_M$ correctly captures the mean marginal structure computed from $\rho_m$  of Fig.~\ref{fig:Mag05}(c), while Fig.~\ref{fig:stat cleaning}(b) shows the computed $\Delta G^{(2)}$.  Although the signal has higher contrast, but for sparsely distributed densities, like for example higher magnification, the relative intensity of the marginal is somewhat similar to the signal, hence a naive subtraction of the marginal image as featured by $\Delta G^{(2)}$ is not enough.
\par Thus, instead of naively subtracting $\rho_M$ , we resorted to structural analysis of both $\rho_M$ and $\rho_m$ using the discrete wavelet transform, which allowed us to isolate and eliminate the embedded  $\rho_M$. 
Discrete \textit{wavelet} based decompositions are known for their ability to sort out spatial frequencies at each location of the image \cite{Wavelet_Vetterli_1992,Wavelet_Strang_1996}. In our case we used  Daubechies wavelets which are widely utilised for robust edge detection in noisy images due to their excellent space-spatial frequency localisation property \cite{Li2023,Li2025}. The corresponding images are only approximates of $\rho_m$ and $\rho_M$, shown in Fig.~\ref{fig:stat cleaning}(c) and (d) respectively, which capture low- to mid-frequency structures. From power spectral density analysis it is seen that approximate $\rho_M$ has a smaller spatial frequency bandwidth than approximate $\rho_m$. Hence, using the spectral band of approximate $\rho_M$, we perform a high-pass Fourier filtering of $\rho_m$ to obtain the form shown in Fig.~\ref{fig:stat cleaning}(e) where the embedded marginal noise is completely removed. 
\par Next, we reconstruct the density $\rho_m$ using inverse wavelet transform as shown in Fig.~\ref{fig:stat cleaning}(f). This is followed by a low-pass Fourier filter and Gaussian smoothening using a Gaussian kernel density estimator that finally leads to the clean density shown in Fig.~\ref{fig:Mag05}(f). 
\section{Experimentally computed density functions for various Magnification}
\label{app:All plots}
The section shows the computed densities and fitted 2D Gaussian for  the rest of the experimental points of Fig.~\ref{fig:Mag variation}(e). They show good fit fidelity for all the magnifications. 
\begin{figure}[!htbp]
    \centering
    \includegraphics[width=1\linewidth,trim = 0.9cm 0.25cm 1.25cm 0.9cm,clip]{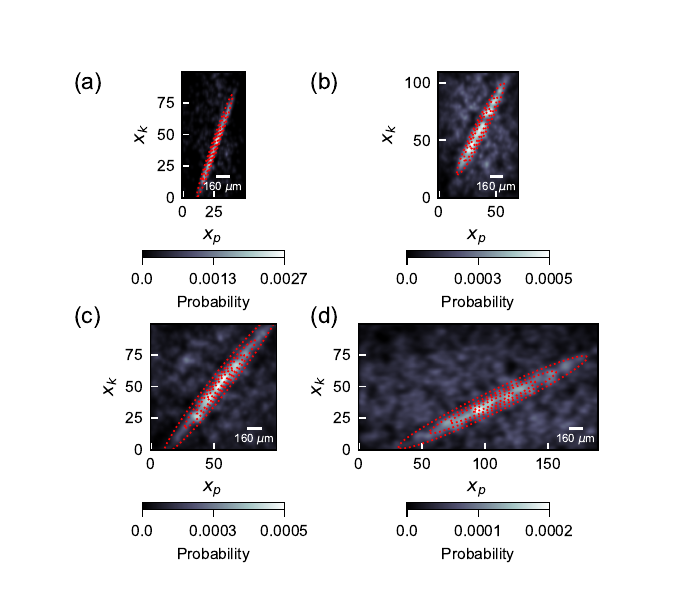}
    \caption{The density plots for the rest of the magnifications $M_m\approx$ (a) $0.5$, (b) $0.75$, (c) $1.25$, (d) $3.0$. These plots are overlaid with fitted 2D Gaussians represented by dotted contours. }
    \label{fig:MagPlotApp}
\end{figure}

\end{document}